\documentclass[prb,twocolumn,superscriptaddress]{revtex4}

\usepackage{amsmath,amssymb,graphicx}
\begin{document}
\title{Spin-Filtering  Multiferroic-Semiconductor Heterojunctions}	
\author{Na Sai}
\affiliation{Department of Physics, The University of Texas at Austin, Texas, 78712}
\author{Jaekwang Lee}
\affiliation{Department of Physics, The University of Texas at Austin, Texas, 78712}
\author{Craig J. Fennie}
\affiliation{Center for Nanoscale Materials, Argonne National Laboratory, Argonne, IL, 60439} 
\author{Alexander A. Demkov}
\affiliation{Department of Physics, The University of Texas at Austin, Texas, 78712}
\date{August 24, 2007}
\begin{abstract}
We report on the structural and electronic properties of the interface between the multiferoic oxide 
YMnO$_3$ and wide band-gap semiconductor GaN studied with the Hubbard-corrected local spin density approximation (LSDA+U) to density-functional theory (DFT). We find that the band offsets at the interface between antiferromagnetically ordered YMnO$_3$ and GaN are different for spin-up and spin-down states. This behavior is due to the spin splitting of the valence band induced by the interface. The energy barrier depends on the relative orientation of the electric polarization with respect to the polarization direction of the GaN substrate suggesting an opportunity to create magnetic tunnel junctions in this materials system.
\end{abstract}
\maketitle

Multiferroics are materials in which ferroelectricity and magnetism coexist in a single phase. 
Efforts have shifted from the question of ``coexistence'', to identifying strategies 
that will increase the coupling between the two orderings. This increased functionality
opens opportunities for novel electrically or magnetically controlled 
devices. One route that is promising for practical applications is to employ multiferroic thin films and multilayered structures, the properties of which can be 
readily manipulated at the nanoscale.\cite{Ramesh07p21,Prellier05p803}
Multiferroic YMnO$_3$\cite{vanAken04p164, Fennie05p100103} is considered an attractive candidate
for use in transistor devices because of its purity from volatiles such as Pb or Bi and 
its moderate dielectric constants.

Much effort has been directed at synthesizing and characterizing thin films of hexagonal 
YMnO$_3$ as a potential gate dielectric for semiconductor devices, e.g.,YMnO$_3$ on 
Si(111)\cite{Fujimura96p1011,Fujimura96p7084,Yoshimura98p414,Yi98p903}, on wurtzite GaN, or on ZnO\cite{Posadas05p171915, Chye06p132903, Bala06p1807}.  
Despite the remarkable progress in synthesis, the role of the interface $-$ strain, chemistry, etc., $-$ on the ferroelectric and magnetic properties of YMnO$_3$ thin films is poorly understood and
difficult to separate experimentally. In this Letter, we apply density-functional theory to 
calculate the electronic structure and band alignment at a realistic YMnO$_3$/GaN interface. We 
demonstrate that interfacial spins behave differently from those in the bulk. This interface
effect leads to a spin splitting in the valence bands giving rise to different band offsets for spin 
up and down states. Intriguingly, the difference in the band offsets depends on the polarization 
direction of YMnO$_3$ relative to that of the polar GaN substrates, suggesting that 
the system could be utilized in spin-filtering tunneling junctions.  

\begin{figure}
\includegraphics*[width=8cm]{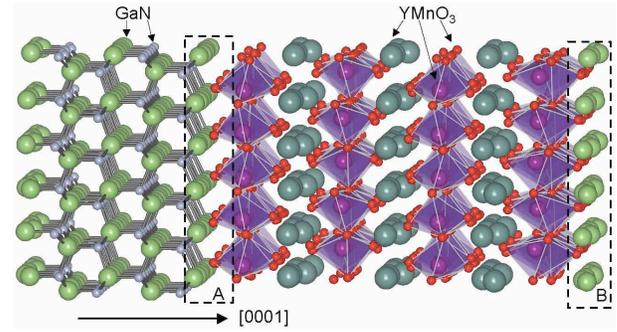}
\caption{The structural model of the  YMnO$_3$-GaN heterojunction. The two inequivalent interfaces viewed along (0001) and (000$\overline{1}$) are defined as $A$ and $B$, respectively.}
\label{str}
\end{figure}

Bulk YMnO$_3$ and wurtzite GaN both have hexagonal symmetry. YMnO$_3$ is antiferromagnetic 
(AFM) and ferroelectric (FE) (space group $P6_3cm$) while GaN is polar but not FE.
X-ray diffraction of YMnO$_3$ thin films deposited on GaN (0001) substrates suggests an in-plane 
rotation of 30$^\circ$ between the unit cell axes of YMnO$_3$ and GaN. For a coherent interface this 
implies that YMnO$_3$ is under a 10\% in-plane compressive 
strain.\cite{Posadas05p171915,Chye06p132903, Bala06p1807} This large epitaxial strain has been 
attributed by Posadas {\it et al.} to the energy gain from the interfacial bond 
formation.\cite{Posadas05p171915} Here, we adopt the experimentally determined interfacial relation 
and build heterostructures composed of two unit cells of YMnO$_3$ and GaN each, with a total of 24 atomic layers (Fig.~\ref{str}). Both YMnO$_3$/GaN interfaces are comprised 
of Ga and apical oxygens, coming from YMnO$_3$ and GaN, respectively. The interfacial oxygens are placed 
above the {\it fcc} sites ({\it i.e.}, the threefold cavity sites) of the Ga surface. The periodic boundary 
condition in our calculations creates two inequivalent Ga-O bonded interfaces between the sequence 
of YMnO$_3$ and GaN, one with oxygens above the Ga (0001) face and another with oxygens above 
the Ga (000$\overline{1}$) face (see interfaces $A$ and $B$ in Fig.\ref{str} respectively).

We perform DFT calculations of YMnO$_3$ within the LSDA+U approximation with $U=6$ eV and $J = 0.9$ eV.\cite{note:VASP} The value of $U$ 
was extracted from experiment (XPS\cite{Kang05p092405}). We obtain a band gap of 
1.46 eV (1.47 eV)\cite{Kang05p092405} and lattice parameters of $a_{\rm YMO}$=6.09\AA\ 
(6.127\AA)\cite{vanAken01p230} and  $c/a$ = 1.86 (1.86) which agree well with 
experiment (shown in parentheses). 
The Mn spins were treated in a frustrated collinear-AFM approximation as the 120$^{\circ}$ noncollinear 
spin structure observed in bulk YMnO$_3$ is beyond the computational capabilities for
the realistic interface that we consider here.
For bulk GaN, we carried out 
LDA calculations and found $a_{\rm GaN}$ = 3.15\AA, $c/a = 1.627$, and $u=0.377$. The LDA 
band gap is 2.1 eV, which is below the true value 3.5 eV. Since the Hubbard correction  
does not provide genuine improvement in GaN, we used a Hubbard U only on the Mn atom. 

To isolate the effect of strain, we first relax bulk YMnO$_3$ under in-plane compressive strain 
of $\sqrt{3}a_{\rm GaN}/a_{\rm YMnO}\sim10\%$. The $c$-axis lattice parameter expands by 
$\sim$7\%, consistent with Ref.~\onlinecite{Posadas05p171915}. The Mn moments are slightly 
reduced $\sim3.6\mu_B$ compared to the unstrained bulk value of $3.77\mu_B$. The distances 
between Mn and the in-plane oxygens Mn$-$O$_P$ shorten to $\sim1.86$\AA, while the distances 
between Mn and the apical oxygens Mn$-$O$_A$ lengthen to $\sim1.9$\AA, compared to 
$\sim2.04$\AA\ and $\sim1.86$\AA, respectively, in the unstrained bulk. Despite the 
significant shortening of the Mn-O$_P$ bonds, no changes occur to the topmost valence bands 
including O$_P$~$p_x$,$p_y$ and Mn $d_{xy}$ and $d_{x^2-y^2}$ orbitals as well as the unoccupied 
$d_{z^2}$ states. The most noticeable change occurs in the O$_A$~$p$ states and the O$_P$~$p_z$ 
states immediately below the O$_P$~$p_{x,y}$ bands, which downshift by $\sim0.6$ eV near the 
$\Gamma$ point, presumably due to strain-induced buckling of MnO$_5$ bipyramids. 

\begin{figure}
\includegraphics*[width=7cm]{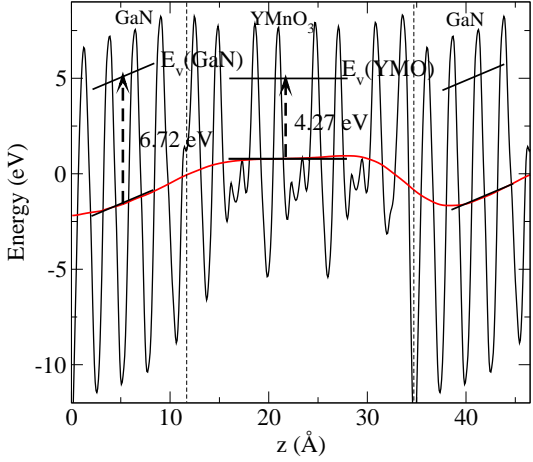}
\caption{The profile of the macroscopically averaged electrostatic potential across the YMnO$_3$-GaN supercell $P_1$. The straight lines are the top of the valence bands measured with respect to the average of the potential in each material, the values are referenced from independent bulk calculations.}   
\label{epot}
\end{figure}

Because wurtzite GaN is polar (a permanent dipole moment is produced by each bilayer of 
Ga$^+$ and N$^-$) we construct two supercells; ``$P_1$'' with the FE polarization in 
YMnO$_3$ pointing along GaN (000${\overline 1}$), and ``$P_2$'' with the FE polarization 
pointing along GaN (0001).  We then relax both supercells fixing the in-plane lattice
constant to $a_{\rm GaN}$.  In either supercell, the vertical distances between the apical 
oxygens and the Ga atoms at the $A$ and $B$ interfaces (see Fig.\ref{str}) are drastically 
different, ranging from 1.1$-$1.2\AA\ at interface  $A$ to 0.2$-$0.6\AA\ at interface $B$. The 
different bonding structures of two interfaces agrees with the behavior of oxygen adatoms 
absorbed on Ga surfaces in the low oxygen coverage (growth) condition of GaN.\cite{Zywietz99p1695} 

\begin{figure}
\includegraphics*[width=8.5cm]{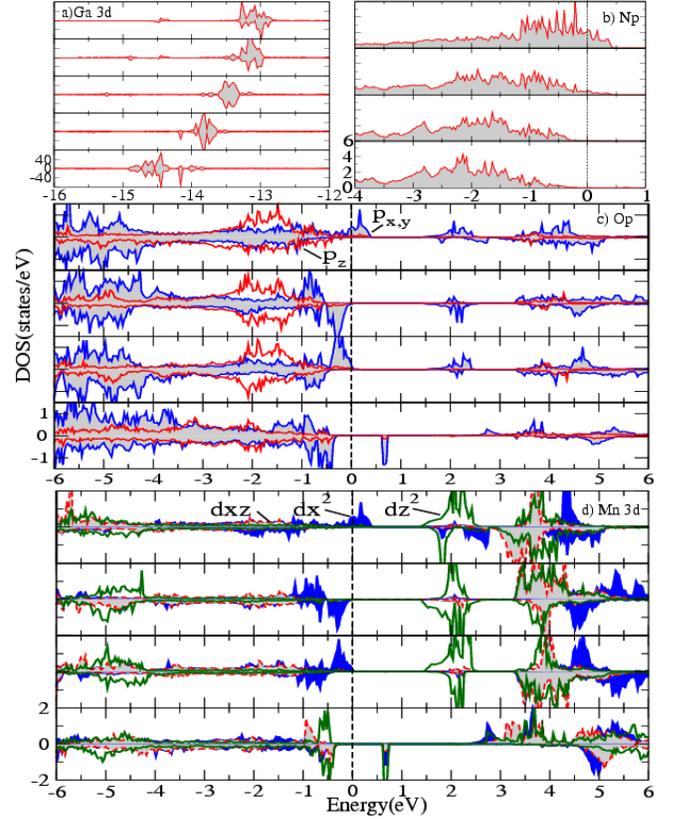}
\caption{Orbital-resolved DOS projected on atomic layers of the YMnO$_3$-GaN supercell shown in Fig.~\ref{str}. In each panel, the top and bottom most graph corresponds to interface $A$ and $B$, respectively. The projected DOS of Mn $d_{xy}$ and $d_{yz}$ (not shown) are nearly degenerate with $d_{x^2}$ and $d_{xz}$ respectively shown in d). The vertical lines mark the Fermi level.}   
\label{dos_atomic}
\end{figure}
Fig.~\ref{epot} shows the macroscopically averaged electrostatic potential the $z$ direction. The polarization field in GaN, evident from the slope of the electrostatic potential, induces charge transfer across the film and induces screening on the opposite sides of the GaN films. As a consequence, the residual field across YMnO$_3$, parallel (antiparallel) to the field in GaN in supercell $P_1$ ($P_2$), is small. Using the effective ionic charges,\cite{vanAken04p164} we estimate the polarizations of YMnO$_3$ in the supercell to be $-9.99\mu$C/cm$^2$ for $P_1$ and $8.43\mu$C/cm$^2$ for $P_2$. These numbers are slightly lower than $14.0\mu$C/cm$^2$ calculated for the constrained bulk ($P\sim 8.8\mu$C/cm$^2$ in unconstrained bulk)\cite{vanAken04p164}, indicating incomplete screening of the polarization charges at the interfaces. The polarization fields in these systems complicate comparison of the band edges directly from the lineup of the average potential as shown in Fig.~\ref{epot} 

To analyze the electronic structure of the supercells we calculate the orbital-resolved, layer-projected density of states (PDOS). In the $P_1$ supercell, there is clearly a band bending of $\sim1$ eV in GaN due to polarization, as illustrated by the deep Ga $3d$ valence states in Fig.~\ref{dos_atomic}$a$. The band bending is also evident in the topmost valence bands N$_p$ in Fig.~\ref{dos_atomic}$b$ that produces an upward bending  at interface $A$ and a downward bending at interface $B$, consistent with the experiments in GaN (0001) films.\cite{Chevtchenko06p2104} Despite this bending, there is no sign of gap closing as the thickness of our GaN films is much below critical.\cite{Fiorentini99p8849} Fig.~\ref{dos_atomic}$c$ and $d$ show the DOS for Mn $3d$ and planar O$_P$ $2p$ states. The interior layers look similar to the bulk layers, suggesting that the interior region converges to the bulk. The hole states at interface $A$ are composed of Mn $d^\uparrow_{xy}$, $d^\uparrow_{x^2-y^2}$, O$_P$ $p^\uparrow_{x,y}$ and a small contribution from O$_A$ $p^\uparrow_{x,y}$. At interface $B$ we observe a down shift in energy of the Mn $3dz^2$ orbitals (located at $\sim$2eV above $E_F$ in the bulk), and a strong overlap with the O$_p$ states. We encounter similar behavior in supercell $P_2$ except that the $E_F$ is $\sim0.2$ eV deeper than in $P1$. The difference can be related to charge compensation at the interfaces: in $P_1$, the polarizations of YMnO$_3$ and GaN create opposite screening charges that partially cancel at the interfaces, while in $P_2$, the polarizations create screening charges of the same sign and thus lead to higher accumulation than in $P_1$.  

In both supercells we observe that the spin moments of Mn at the YMnO$_3$/GaN interfaces 
deviate from their bulk values. On average the moment of Mn reduces to $\sim3.3\mu_B$ 
at interface $A$ and increases to $\sim4.2\mu_B$ at interface $B$,  producing a small net
magnetization of $1.5\mu_B$ per supercell. This can be understood from the DOS analysis which 
shows a depletion of charge on the Mn $d_{xy}$ and $d_{x^2}$ at interface $A$ and accumulation of 
charges on the Mn $d_{z^2}$ orbitals at interface $B$.  It is certainly possible that the precise route to the observed ferrimagnetic moment depends on the collinear approximation used in this study, but the basic physics of spin manipulation at the YMnO$_3$/GaN interface should be qualitatively similar if noncollinear spins are considered. 

\begin{figure}
\includegraphics*[width=8cm]{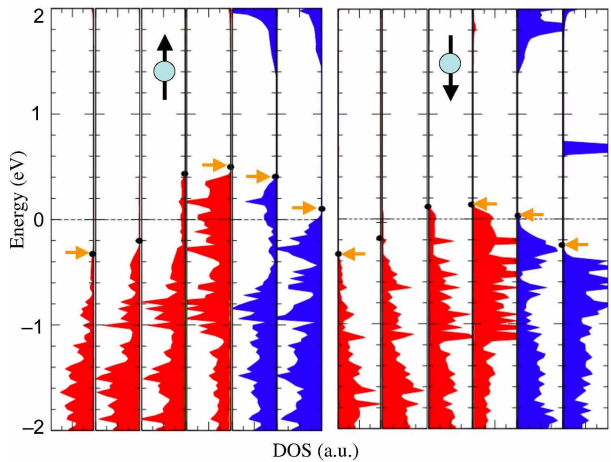}
\caption{Projected DOS of the YMnO$_3$-GaN $P_1$ supercell into 4 bilayers of GaN (red) and 2 unit-cells of YMnO$_3$ (blue). The left and right panels are DOS for spin-up and spin-down components. The black dots mark the valence band edge and the arrows mark the interface band edge.}   
\label{dos_unitcell}
\end{figure}
To calculate the band offset at the YMnO$_3$/GaN interface, we project the spin-resolved DOS of the supercell onto unit cells of YMnO$_3$ and bilayers of the GaN as shown in Fig.~\ref{dos_unitcell}, and calculated the band offsets for both the spin-up and spin-down components. In the case of $P_1$, the valence band offset (VBO) is $-0.05$ eV for the spin-up component and $-0.1$ eV for the spin-down component at interface $A$, whereas a VBO is $+0.35$ eV for the spin-up and $+0.15$ eV for the spin-down at interface $B$ (``$+$'' and ``$-$'' represent upward and downward offsets going from GaN to YMnO$_3$). The band offset at both interfaces shows spin dependence. In particular, the spin-up electrons experience a potential higher by 0.2 eV than the spin-down electrons across interface $B$. This suggests that electrons of different spin directions experience different electrostatic scattering probabilities and that {\it spin-dependent barriers} can be produced in AFM multiferroic/semiconductor heterostructures. This effect is associated with spin splitting at the top of the valence bands. Recently, BiMnO$_3$ tunnel barriers were shown to be potential spin filters in magnetic tunnel junctions.\cite{Gajek07p296} But FM multiferroics such as BiMnO$_3$ are rare. Thus our observation of spin-dependent barriers in AFM multiferroic/semiconductor junctions might enable the use of a larger group of multiferroics with AFM orderings as spin filtering devices. 

In the supercell $P_2$ (not shown), we find a VBO of $0.1$ eV for the spin-up and $-0.1$ eV for the spin-down at interface $A$, similar to that of $P_1$. At interfaces $B$, the VBO is $+0.2$ eV for spin-up and $+0.4$ eV for spin-down, slightly higher in amplitude than that of $P_1$. It is noteworthy that the offset of the $P_2$ structure is reversed for up and down spins from that in $P_1$ at interface B, which suggests a possibility of controlling the spin scattering by a reversal of the ferroelectric polarization in YMnO$_3$. A small energy difference of $\sim 20$ meV is found in bulk YMO$_3$ when the spins are switched from the collinear to noncollinear in-plane configurations.\cite{Fiebig05p883} This is much less than the barrier differences we found between the up and down spins. Thus the magnetic anisotropy should not influence the spin-dependent effect we observe. 

In conclusion, we report the details of the electronic structure of hexagonal YMnO$_3$-GaN heterojunctions. We consider two inequivalent Ga-O terminated interfaces that can be found in YMnO$_3$ films grown on (0001) and (000${\overline 1}$) oriented GaN substrates and two possible orientations of the YMnO$_3$ polarization with respect to that of the GaN substrate. We find different band offsets for spin-up and spin-down components, with a larger variance at the (000${\overline 1}$) interface. The spin-dependent interface barriers suggest that these heterostructures may be applicable in spin filtering tunneling devices. Our results are relevant not only to YMnO films but also to other multiferroic thin films with coexisting antiferromagnetic and ferroelectric structures.    
  
This work is supported by the Office of Naval Research under grant N000 14-06-1-0362 and Texas Advanced Computing Center. 


\begin{thebibliography}{10}
\bibitem{note:VASP} We have performed the LDA+U calculations using the Vienna Ab initio Simulation Package (VASP) with projector augmented wave potentials, a cutoff energy of 500 eV, and a Monkhorst-Pack $\Gamma$-centered $4\times4\times2$ $k$-points mesh for the bulk and supercell calculations. The supercell DOS was generated with a denser $6\times6\times4$ mesh. The forces from ionic relaxation of the supercells are  $<8$meV/\AA.
\end{thebibliography}

\end{document}